\documentclass[twocolumn]{revtex4}

\usepackage{float}
\usepackage{graphicx}
\usepackage{wrapfig}

\begin{document}

\title{Modelling the Spatial Dynamics of Culture Spreading in the Presence of Cultural Strongholds}

\author{Ludvig Lizana}
\affiliation{Niels Bohr Institute, Blegdamsvej 17, DK-2100, Copenhagen, Denmark}
\author{Namiko Mitarai}
\affiliation{Niels Bohr Institute, Blegdamsvej 17, DK-2100, Copenhagen, Denmark}
\author{Hiizu Nakanishi}
\affiliation{Department of Physics, Kyushu University 33, 812-8581 Fukuoka, Japan}
\author{Kim Sneppen}
\affiliation{Niels Bohr Institute, Blegdamsvej 17, DK-2100, Copenhagen, Denmark}
\date{\today}

%
%

\begin{abstract}
Cultural competition has throughout our history shaped and reshaped the geography of boundaries between humans. Language and culture are intimately connected and linguists often use distinctive keywords to quantify the dynamics of information spreading in societies harbouring strong culture centres. One prominent example, which is addressed here, is Kyoto's historical impact on Japanese culture. We construct a first minimal model, based on shared properties of linguistic maps, to address the interplay between information flow and geography.  In particular, we show that spreading of information over Japan in the pre-modern time can be described as a Eden growth process, with noise levels corresponding to coherent spatial patches of sizes given by a single days walk, and with patch-to-patch communication time comparable to the time between human generations. 
\end{abstract}

\maketitle

It is generally understood in historical linguistics that geolinguistic diffusion, the process by which linguistic features spread geographically from one dialect or language to another,  plays a central role in the evolution of languages \cite{boberg2000geolinguistic, labov2003pursuing, labov2007transmission}.  Origins of linguistic changes are plentiful where  societal changes and movements are of pivotal importance. The dynamics and causes of linguistic change provide therefore important clues to the historical developments of, as well as interplay between, societies and civilisations \cite{gray2009language, welsch1992language}.
It has long been observed that linguistic features, just like innovations, spread outward from an originating centre \cite{bailey2008some}.  Spatial patterns are, however, rather ambiguous. Some cases \cite{trudgill1974linguistic} bear evidence of a hierarchal diffusion process where e.g. dialect changes propagate in a cascade-like manner from larger to smaller cities. Other cases 
\cite{boberg2000geolinguistic} show an isotropic geographic distribution where linguistic features spread as a wave front among adjacent speech communities. A unifying theme is, however, that recent changes are found close to the~source.

In this paper we study the dynamics of culture spreading around strong culture centres. As a proxy for the spreading of cultural traits we use the spreading of words and put forward a model concerning the spatial dynamics of competing wave fronts where the key feature is that new words are more prone to be adopted than old. This view contrasts most models dealing with information spreading, whether it concerns innovations \cite{rogers1995diffusion}, opinions \cite{liggett1999stochastic} or linguistic traits \cite{trudgill1974linguistic}, where two different pieces of information are treated on an equal footing. Our take on the problem is different. It is based on the fundamental property that {\it the value of information decays with time} \cite{stiglitz1981credit}. Previous studies \cite{rosvall2006self, lizana2010time} have established that an ongoing replacement of old information by new has major consequences for its spatio-temporal dynamics.

As a case study we consider the geographic distribution of words over Japan. Carful analysis of linguistic maps \cite{lingmap} has unveiled that words in many cases are arranged in concentric ring-like patterns with Kyoto, Japan's ancient capital, as focal point.  This was first realised by Kunio Yanagita, a famous Japanese folklorist, who studied the distribution of the word for snail ({\it kagyu}) \cite{Yanagita}. He also found that the same old variants were used in the southern and northern parts of the country but not in the middle. This lead him to formulate the "periphery propagation theory of dialects" ({\it Hougen Shuken-ron}) in which he visualised waves of new words emerging in Kyoto which  spread radially outwards. 

\begin{figure*}[]
\begin{center}
\includegraphics[width=18cm]{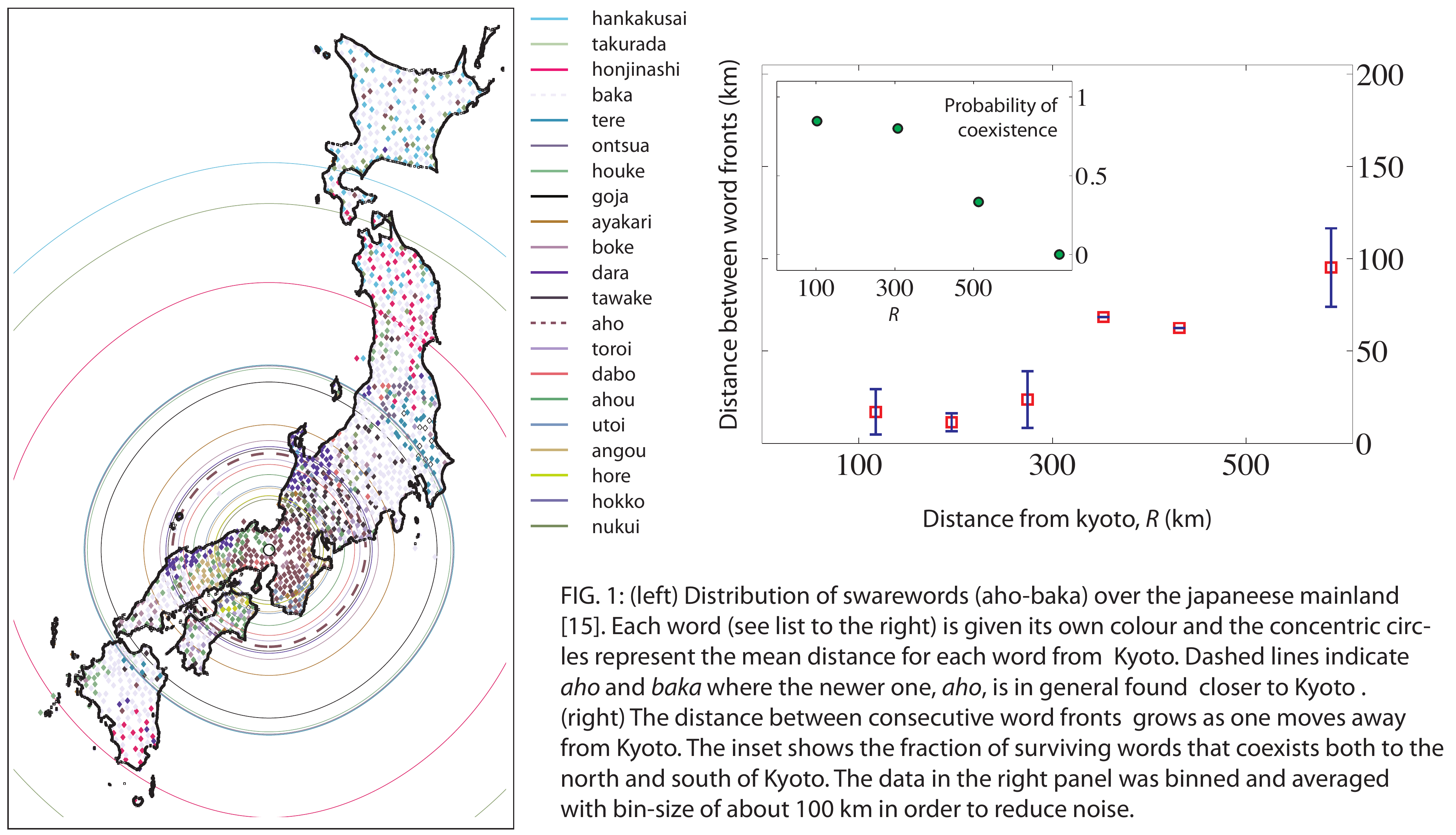}
\caption{}
\label{fig:ahobaka}
\end{center}
\end{figure*}

The most beautiful example of Yanagita's theory is the distribution of swearwords. The Japanese are not known for their frequent use of swearwords, but if you nevertheless are cursed at by someone with {\it baka} ($\sim$ stupid person), the one you are having trouble with is probably from Tokyo.  If you instead hear {\it aho} ($\sim$ dumb), he or she is most likely from the Kyoto-Osaka area. The confrontation between these two swearwords is so clear it is considered by the people as a part of the competition between the two major cultural centres. The result of a comprehensive survey of the different variants of  {\it aho-baka} \cite{ahobaka} is displayed to the left in Fig. \ref{fig:ahobaka}; The concentric patterns centred around Kyoto are unmistakable. {\it Aho} and {\it baka} are indicated by dashed lines and Tokyo is a part of the circular area where {\it baka}, not {\it aho}, is being used. {\it Baka} used to exist in Kyoto in the past but has been overrun by the newer {\it aho}. Other noticeable features are that the  area of the word patches grow with increasing distance from Kyoto (Fig. \ref{fig:ahobaka}, right), and that only a fraction of the words are found both to the north and south of Kyoto (e.g. {\it goja} is only found to the north). These observations are well described by our model.

Our model is defined on a two dimensional lattice on which words, after being coined in the culture centre, spread. In order to capture the ongoing adoption and subsequent communication of new words originating with a given frequency $f_{\rm word}$, at each time step a word is replicated and passed on to a neighbouring randomly chosen lattice site. If it is sent to a location inhabited by an older version, the new one is adopted. But, if it is transmitted to a place where an even newer variant exists, the older word is ignored; new concepts always overrules old. We point out that our model captures the way in which new words invade new territories and not their coexistence. 
We have implemented our model in an interactive on-line  java applet \cite{applet} which can be used by anyone who wishes to explore the properties of our model.

\begin{figure*}

\includegraphics[width = 18cm]{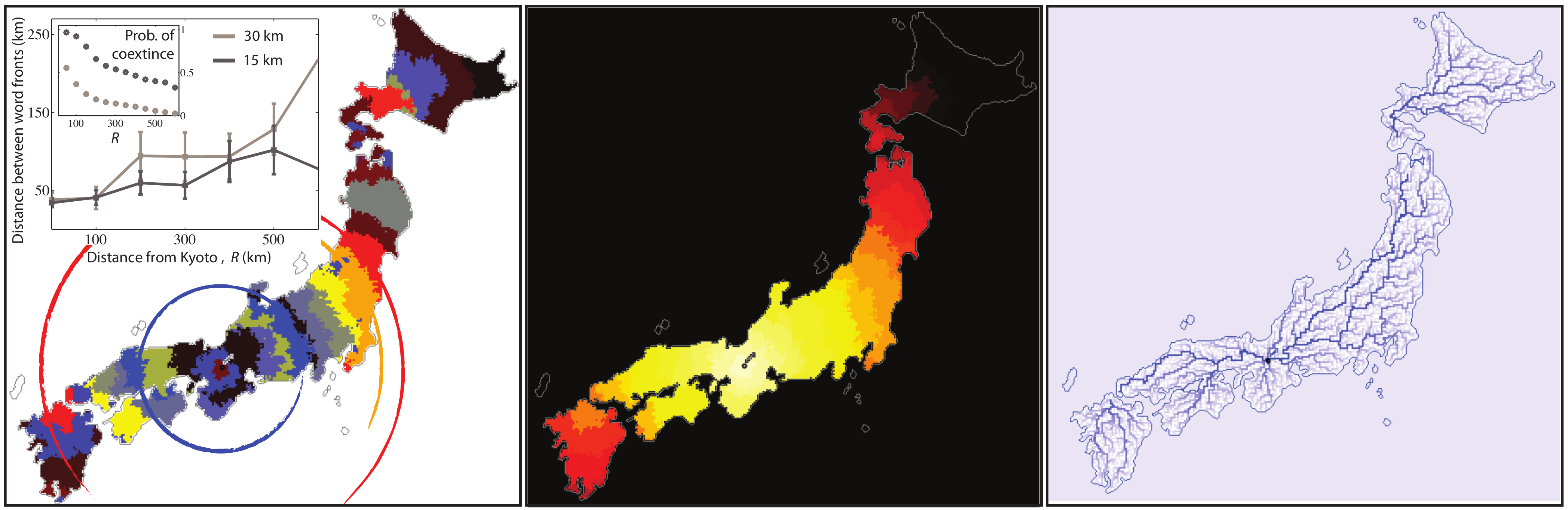}

\caption{Snapshots of a simulation showing the spatial dynamics of word spreading  over the Japanese mainland. 
(left panel)
Ongoing spreading where each colour represent a different word. Blue and red circles show two examples where the same word form is found symmetrically on either side of Kyoto. The graph in the upper left corner shows the mean distance between two adjacent fronts (averaged over many runs) as a function of distance from Kyoto. The orange broken circle belongs to a word which only is present at Kyoto's north east side. The probability that a surviving word coexists on both sides decay with distance away from Kyoto in a way shown in the inset. We investigated the behaviour when the spatial resolution (width of lattice site) was set to 15 and 30 km, respectively.
(middle panel)
Age landscape illustrating how older and older words (yellow to dark red colouring) are encountered as one moves further away from Kyoto.
(right panel)
River landscape showing the paths the new words took as they left Kyoto.
In order to improve figure quality, a lattice spacing of $\Delta = 5$ km was used.
}
\label{fig:japan}
\end{figure*}

Figure \ref{fig:japan} shows three snapshots of our java applet. In the left panel each new word is given its own random colour and Kyoto is marked in black. Our model gives rise to patterns of concentric coherent patches penetrating the landscape moving outward from the source. In particular, notice how the same colour is found on either side of Kyoto without being present in the middle, just like in the real data (Fig. \ref{fig:ahobaka}). The blue and red circles highlight two examples. If we calculate the corresponding radii for all colours in the landscape and average over many landscapes, the mean distance between two consecutive words increase as a function of distance from Kyoto as is shown in the in the left panel. This result obviously hangs on the frequency of new words $f_{\rm word}$ from Kyoto as well as the coarse graining of space. The graph therefore depicts two cases where we used lattice spacings of $\Delta = 15$ km and $\Delta = 30$ km where the word frequency was adjusted in each case such that about $20$ words could be distinguished simultaneously on Honshu island, as in Fig. \ref{fig:ahobaka}. Based on the estimate $v_{\rm word}=1$ km/year \cite{wordspeed} we find from our model that new words are being coined in Kyoto on average every 30th year for $\Delta = 15$ km and every 60th year for $\Delta = 30$ km. 
Another feature is that some words are only found on one side of Kyoto, and that this correlates with distance. One example is indicated by the orange arc. We thus measured the fraction of surviving words that existed on both sides as a function of distance and the result is shown in the inset of the graph. For the crude coarsening ($\Delta = 30$ km), the probability decays quite rapidly and at $R=300$ km (southern tip of Honshu) as little as 14\% of the surviving words coexist on both the north and south side of Kyoto,
while the corresponding number is  $54\%$ for the finer lattice  ($\Delta = 15$ km). These results, combined with that the increase in the width of the colour patches for the $\Delta = 15$ km case match the real data better, leads us to conclude that this constitutes a reasonable level of spatial coarsening for our~system. The middle panel in Fig. \ref{fig:japan} shows the word age distribution at the same time point as portrayed in the left panel; There is a clear age gradient (light to dark) from Kyoto towards the north and south parts of the country.

In addition to the geographical distribution of words our model predicts along which routes the words travelled. To do this we at each lattice point 
remember not only the age of the latest word, but also where it came from.
This means that starting from any point in the plane we can follow information pointers downstream and eventually reach all the way back to the source. If the landscape is frozen at some time point and the paths from all points in the plane are mapped out 
we obtain an ``information river network" that is shown to the right in 
Fig. \ref{fig:japan}.  River depth, symbolised by light to dark blue colouring, is proportional to the number of reachable upstream lattice points. The predicted river network is self-similar and obeys similar scaling relations as river networks of flowing water (but with different exponents).  
The real ``information rivers" will in addition be influenced by 
other factors, such as mountain ranges and variations in population density.

Our model of replicating information is based on the idea that new information always out beats old. This is a simple principle which, 
of course, is not always true. With this in mind, it is interesting to see the consequences if old information sometimes can win over new, say with a probability $p$. Running the model with this modification effectively reduces the spreading speed of words, and leads to fragmentation and increased roughness of the word patches (see Fig. \ref{fig:figS1} in appendix \ref{App:A} for the $p=0.8$ case).  In the limit $p \rightarrow 1$, where old information win over new just as often as the other way around (i.e. unbiased diffusion of words), new words have major difficulties leaving Kyoto and the ring-like structure of the word landscape is lost.

The density of word fronts as well as the probability of two words being present on either side of Kyoto, is connected to the annihilation of old words by new ones catching up from behind. In order to increase our understanding of the dynamics of the problem, we ran our model on a simpler system than the Japanese mainland, namely an elongated rectangular lattice with width $L$. Placing a line source at the base, our simulations demonstrate (see appendix \ref{App:B}) that the spacing between two word fronts increase as $z^{0.5}$, for large distances $z$ away from the source. 
%
In one dimension our model maps exactly onto the limiting behaviour of the reaction-diffusion problem $A+B\rightarrow A$ \cite{evans2002nonequilibrium}, for which the density decays with time $t$ as $\rho(t)\sim t^{-0.5}$: relating $z$ to the constant average speed of the surface $v_{\rm surface}$ via  $z=v_{\rm surface} t$, gives $\rho(z)\sim z^{-0.5}$. The mapping is admissible also in two dimensions when the motion of one interface can be represented by a single coordinate (when correlations extends over the whole boundary). 
%

Concentric wave patterns of lingiustic traits surrounding cultural strongholds is by no means unique for Japan. Already in 1872 the German linguist Johannes Schmidt discussed a wave theory (Wellentheorie) for how changes propagate in a speech area \cite{schmidt1871geschichte}. Similar ideas have also emerged in spatial economics \cite{von1966isolated}.  Our model of information spreading also  bears resemblance to other growth models in biology and physics. The average propagation of a new word into the background of an old is similar to a discrete version the wave-like front described by Fishers's equation for bacterial growth \cite{fisher1937wave}. Our model also incorporates stochastic properties of growing interfaces of new information that mimic Eden surface growth \cite{eden1961} in which unoccupied perimeter sites of a growing cluster are filled randomly with probabilities proportional to the number of occupied nearest neighbours. 

The view of an ongoing replacement of "new" with "old" differentiates our work from other growth models where focus has been on one single front \cite{eden1961}, or on the mutual exclusion of bacterial strains
growing into the same free territory \cite{kimura1964stepping, saito1995critical, korolev2010genetic}. By opening for an ongoing 
replacement of one culture with another, 
our model could be generalised to more complicated spatial battles in living systems. Already when considering the spreading of simple words, it is in fact not all words which show regular circular wave patterns. Some patterns are fragmented which presumably reflects a competitive dynamics where the difference between new and old is close to neutral. The main feature of our model is the interplay between coherence within the culture and the ability to transmit information. If we deformed the space on which the information spreads by adding shortcuts, such as roads, our model predicts enhanced cultural coherence. When Romans conquered Europe, they instantly build roads which served as communications lines that kept the provinces culturally coherent with Rome \cite{laurence1999roads}. Likewise to our scheme, the battle for cultural dominance will be ruled by information flow.

We acknowledge the Knut and Alice Wallenberg foundation and the Danish National Research Foundation for financial support, Hirofumi Aoki and Yoshizo Itabashi for letting us know some of the basic literature of Japanese linguistics, and  Martin Rosvall for useful comments.

%
%

\appendix

\section{Dynamics when old information sometimes beats new}
\label{App:A}

\begin{figure}
\begin{center}
\includegraphics[width=\columnwidth]{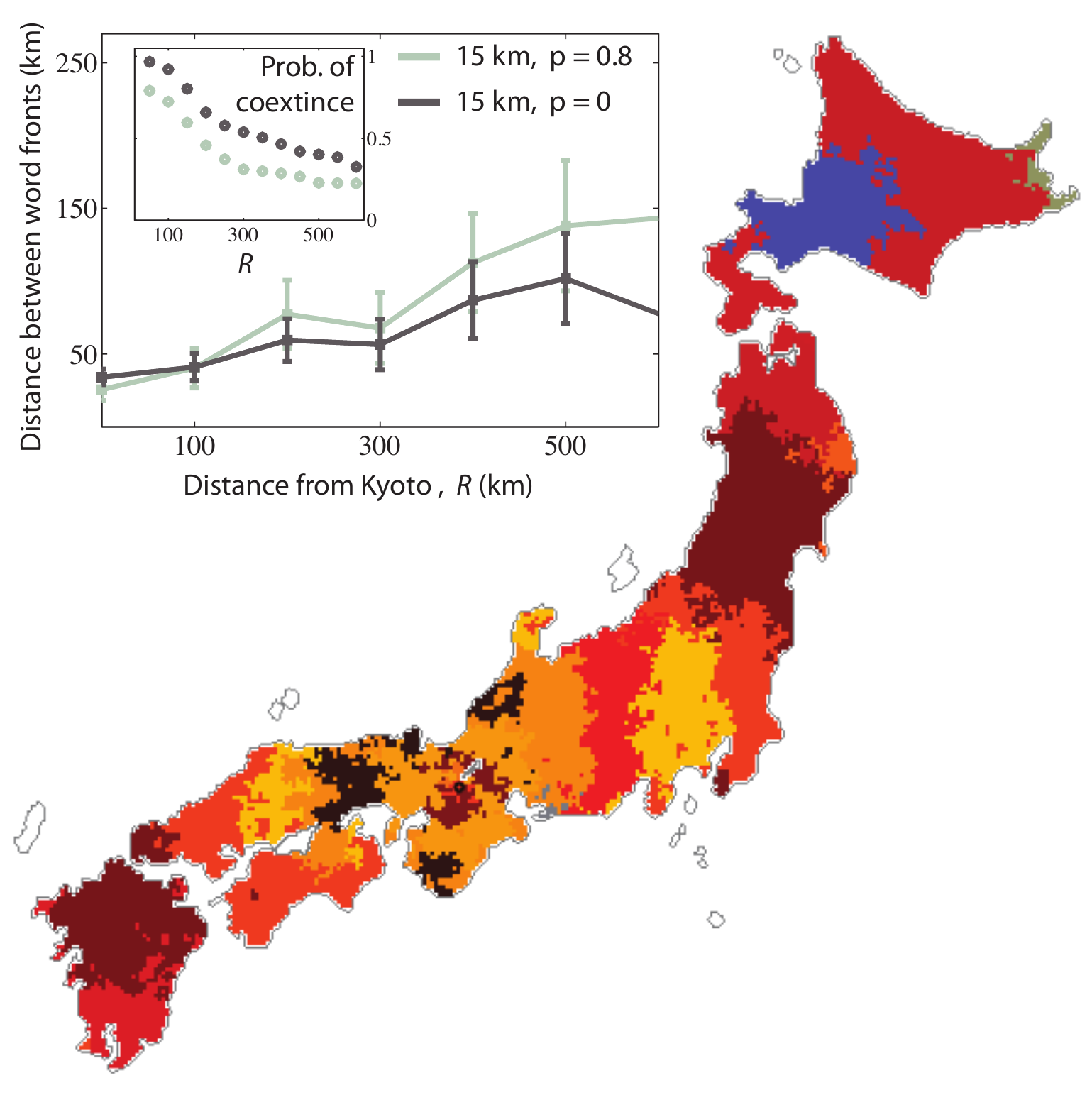}
\caption{Spatial dynamics of words where old words overtake new in 80\% of the cases. Each colour represent a different word.  The graph to the upper left shows how the distance between two adjacent word fronts grow with distance from Kyoto compared to the original model $p=0$ for the coarsening $\Delta = 15$ km.  The probability that the same word variant exists on both sides of Kyoto is depicted in the inset. In order to improve the quality of the word landscape we used a lattice spacing of $\Delta = 5$ km.}
\label{fig:figS1}
\end{center}
\end{figure}

Our model is based on the simple principle that new information is considered more valuable than old. This is indeed a simplification and  it is interesting to see how the model behaves when this condition is relaxed. We thus introduce a parameter $p$ which is the probability that old information can overrule new (new words still, however, always wins over old). For $p=0$ we recover the original model while $p=1$ is the case when new and old information are considered equal. Figure \ref{fig:figS1} shows the case when $p=0.8$ and is analogous to the left panel in Fig. 2 in the manuscript. The concentric word distribution is still visible but the interfaces between the word patches are rougher compared the $p=0$ case, especially close to the centre. Finite values of $p$ thus increase the noise level in the system which also can be seen in the graph in the upper left corner. The speed of the growing interfaces of course decrease with increasing $p$.

\section{Dynamics of the model on a rectangular lattice}
\label{App:B}

\begin{figure}
\begin{center}
\includegraphics[width=\columnwidth]{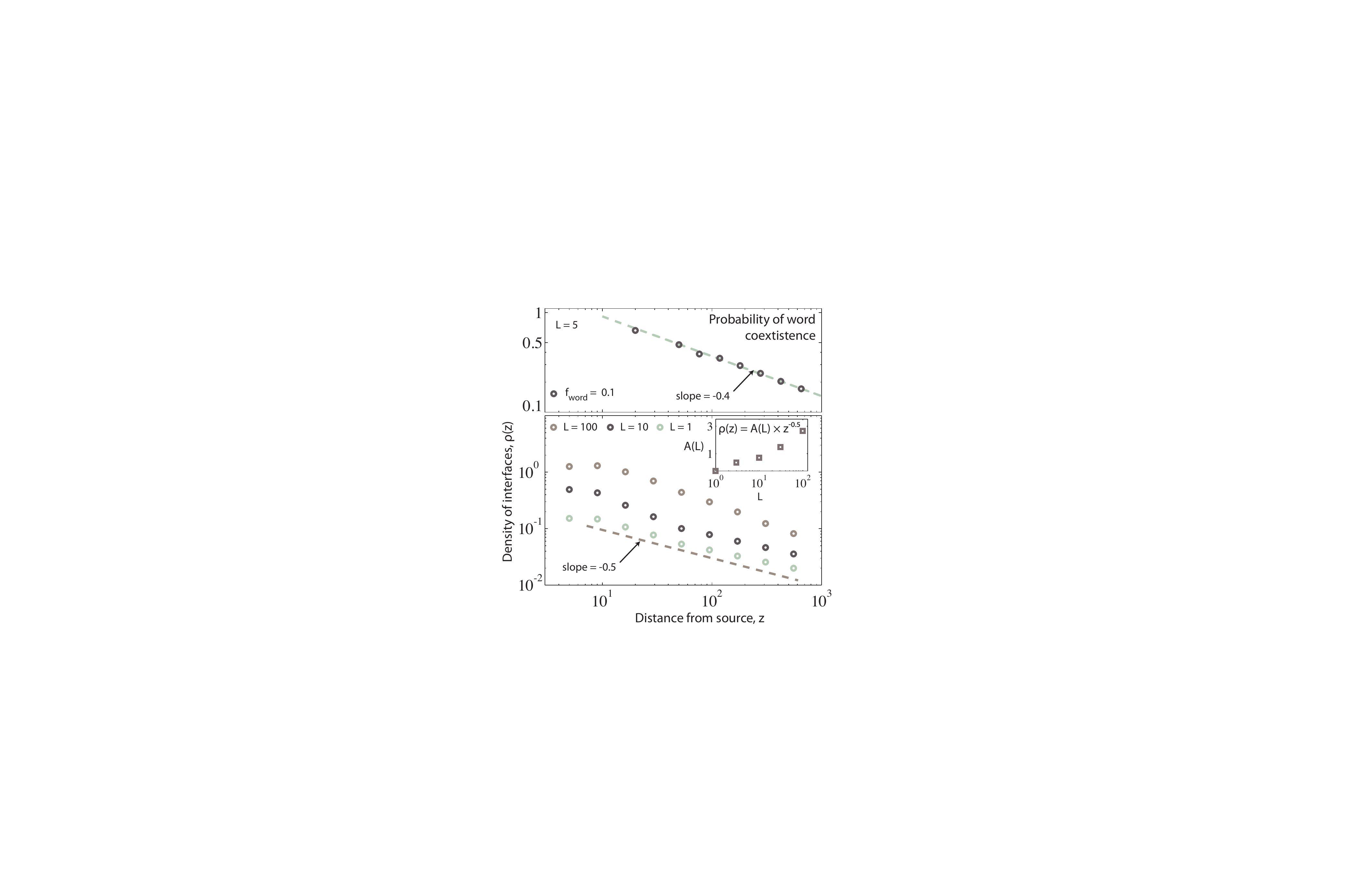}
\caption{Simulation results from a rectangular lattice with width $L$, replication rate $k_R$ and frequency of new words $f_{\rm word}$. 
(bottom) Decay in surface density  $\rho(z)$ as a function of distance $z$ away from the source for three different system sizes. An average is taken over 1000 runs and $f_{\rm word}/k_R=1$.
(inset) System size dependence on the pre-factor $A(L)$.
(top) Probability that a word is present on both sides of the source as a function of distance $z$.
}
\label{fig:figS2}
\end{center}
\end{figure}

In order to improve our understanding of the model, we performed simulations on a rectangular lattice with with width $L$ (which was chosen to be smaller than its height).  New words are introduced along a line source at the base of the lattice at the frequency $f_{\rm word}$ and an invasion attempt to a neighbouring lattice site occurs with rate $k_R$. The results are depicted in Fig. \ref{fig:figS2}.
The lower panel shows that the interface density $\rho(z)$ at large distances $z$ from the source is well described by $\rho(z)\simeq A(L) \times z^{-0.5}$, i.e. the distance between two consecutive word fronts grows as $z^{0.5}$. This holds in one dimension ($L=1$) as well as in two ($L=10$ and $L=100$) dimensions. The dependence of system size  on the pre-factor $A(L)$ in log-log scale is shown in the inset.

The probability that two words exists on both sides of the source can also be quantified in this simple setting if we place the  line source in the middle of the system. The likelihood of coexistence decays with $z$ as is shown in the top panel (for $L=5$) for $f_{\rm word}/k_R=0.01$. The linear fit indicates that the probability decays as $z^{-0.4}$.

%

%
%


\end{document}